\documentclass[aps,twocolumn,pra,showpacs,floatfix]{revtex4}
\usepackage{epsfig}
\usepackage{graphicx}
\usepackage{dcolumn}
\usepackage{amssymb,amsmath}

\begin{document}

\newcommand{\Z}{Z_{\rm eff}}
\newcommand{\Zr}{\Z^{(\rm res)}}
\newcommand{\eps}{\varepsilon}
\newcommand{\eb}{\varepsilon _b}

\title{Relativistic coupled-cluster single-double calculations of positron-atom
bound states}

\author{V. A. Dzuba$^1$, V. V. Flambaum$^1$, G. F. Gribakin$^2$,
and C. Harabati$^1$}

%\author{V. A. Dzuba, V. V. Flambaum and C. Harabati}
\affiliation{$^1$School of Physics, University of New South Wales, Sydney 2052,
Australia}
%\author{G. F. Gribakin}
\affiliation{$^2$Department of Applied Mathematics and Theoretical Physics,
Queen's University, Belfast BT7 1NN, Northern Ireland, UK}

\begin{abstract}
Relativistic coupled-cluster single-double approximation is used to
calculate positron-atom bound states. The method is tested on
closed-shell atoms such as Be, Mg, Ca, Zn, Cd, and Hg where a number of accurate
calculations is available. It is then used to calculate positron binding
energies for a range of open-shell transition metal atoms from Sc to
Cu, from Y to Pd, and from Lu to Pt.
%Fe, Co, Ni, Tc, Rh, Os, Ir, and Pt.
These systems possess
Feshbach resonances, which can be used to search for positron-atom binding
experimentally through resonant annihilation or scattering.
\end{abstract}


\maketitle

\section{Introduction}

In this paper we apply the relativistic coupled-cluster single-double (SD)
approximation to calculate positron binding energies for a number of atoms,
including open-shell transition metal atoms. Positron-atom bound states are
characterized by strong electron-positron correlation effects. These
effects for the positron bound states are even stronger than for their
electron counterparts,
i.e., the atomic negative ions. This is due to the electron-positron
{\em attraction} and, in particular, the large role of virtual positronium (Ps)
formation~\cite{Massey64,Amusia76,Dzuba93}. This makes calculations of
positron-atom bound states a challenging theoretical problem. Using the SD
method allows us to include all the main correlation effects for atoms with
several electrons in the open valence shell, which has not been done before.

The existence of positron-atom bound states was predicted by many-body theory
calculations~\cite{Dzuba95} and verified variationally~\cite{RM97,SC98} more
than a decade ago. Since then positron binding energies have been calculated
for many ground-state and excited atoms: He~$2^3$S, Li, Be, Be~$2^3$P, Na, Mg,
Ca, Cu, Zn, Sr, Ag, Cd and Hg (see,
Refs.~\cite{DFG99,DFH00,MBR02,BM06,BM07,MZB08} and references therein). There
are strong indications that many more atoms, possibly over a quarter of the
whole periodic table, should be able to bind
positrons~\cite{Dzuba95,MBR02,DFG10,Babikov}.
Such conclusions are based on the analysis of the atomic ionization potentials
and dipole polarizabilities, which are relevant for positron binding, and on
the calculations of positron binding to a ``model alkali atom'' \cite{MBR99}
(see also review on positron compounds \cite{Sch09}). In spite of this wealth
of predictions, experimental verification of positron binding to neutral atoms
is still lacking. 

Recently we proposed (Ref. \cite{DFG10}) that positron binding to many
open-shell atoms could be studied experimentally by measuring resonant
positron-atom annihilation. Such resonant annihilation should be similar
to that observed for positrons in many polyatomic molecular species
\cite{GYS10}. In this process the incident positron is captured into
the bound state with the target, with the excess energy being transferred
(in the case of molecules) to vibrations. Since the vibrational motion
of the molecules is quantized, these transitions can only take place at
specific positron energies, which means that they have a {\em resonant}
character. These energies correspond to vibrational Feshbach resonances
of the positron-molecule complex \cite{GYS10,Gr00,Gr01}. The majority of
the resonances observed are associated with individual vibrational modes of
the molecule. The negative energy of the positron bound state $\epsilon _0$
is then related the downshift of the resonance energy,
\begin{equation}\label{eq:enu}
\eps _\nu =\omega _\nu +\epsilon _0,
\end{equation}
with respect to the energy $\omega _\nu $ of the vibrational excitation
\cite{GBS02,BGS03}. Hence, by observing the resonances, the positron binding
energy $\eb =|\epsilon _0|$ can be found. In this way binding energies for
over sixty polyatomic species have been determined
\cite{DYS09,DGS10,JDGNS12,DJGNS12} by measuring positron annihilation using a
high-resolution, tunable, trap-based positron beam \cite{GKG97}.
In our previous paper~\cite{DFG10} we suggested that one can search for a
similar effect in atoms. In this case the resonances will be associated with
low-lying electronic excitations. These can be found in open-shell atoms,
where they often have the same configuration as the ground state. If the
positron can bind to such an atom in the ground state, then it is likely to
bind to its excited state as well. One can then consider the following process,
\begin{equation}
A + e^+ \rightarrow A^*e^+ \rightarrow A^+ + 2\gamma .
\label{eq:Ae+}
\end{equation}
Here the positron first loses energy by exciting the atom and becoming
trapped in the bound state with the excited atom. It then annihilates with
one of the electrons, and the resulting gamma quanta can be detected. 
The first step of the process (\ref{eq:Ae+}) is obviously reversible. Hence,
to estimate the efficiency of resonant annihilation one needs to evaluate the
rates of both positron annihilation and autodetachment~\cite{DFG10}.

To prove the feasibility of such process one also needs to estimate
the positron binding energies to open-shell atoms. In Ref.~\cite{DFG10} we used
a simple approach in which the second-order correlation potential of the
positron-atom interaction was scaled up to account for the effect of
higher-order electron-positron correlations (i.e., virtual Ps formation). 
It was assumed that the scaling factor is the same for all atoms, and
its value was chosen by fitting the results for those atoms where accurate
positron binding calculations are available. Open shells were treated
by using fractional occupation numbers in the standard expressions for a
closed-shell system. This approximation looks
reasonable in the positron-atom problem, because the positron is
not affected by the Pauli principle. The positron-atom interaction
has, thus, no direct sensitivity to the valence shell being open or closed
(unlike the electron-atom interaction in the negative ion problem).

In the present paper we use a more sophisticated fully {\em ab initio}
approach in which the strong electron-positron correlations are included
explicitly through the use of an all-order technique. 

Accurate treatment of the strong electron-positron correlations is the
main challenge in the calculations of positron-atom interaction. It calls
for the use of nonperturbative approaches. For systems with few active
electrons accurate results were obtained using stochastic variation method
\cite{RM97,MBR02}. The most obvious choice for a generic many-electron atom
is the configuration interaction (CI) technique. It was successfully employed
in a number of previous calculations for atoms with one or two electrons in the
valence $s$ shell (see, e.g., \cite{DFG99,DFH00,Bromley02a,Bromley02b}).
However, it becomes
too complicated for more than two valence electrons and, in particular, for
atoms with open $d$-shells, which we want to consider in present work. Another
suitable all-order technique is the couple-cluster approach. In this approach
the interparticle interaction is included to all orders via an iterative
procedure. The corresponding subset of terms includes the so-called
{\em ladder} diagrams~\cite{ladder}. This class of diagrams is very
important in the positron-atom problem, as it describes the effect of
virtual Ps formation. Summation of the electron-positron ladder-diagram
series was performed earlier by solving a linear matrix equation for the
electron-positron vertex function for hydrogen \cite{GL04}, noble-gas
atoms~\cite{Ludlow04}, and halogen negative ions~\cite{LG10}.

The coupled-cluster approach in its single-double (SD) approximation has been
widely used for atoms and ions with one external electron above closed shells
(see, e.g., \cite{SD,BJS91,E3extra,SJD99,DJ07}). It is relatively easy to
modify the corresponding equations for the case of a positron interacting with a
closed-shell atom. We will do this in the next section. 

To test the method we first apply it to positron binding to closed-shell
atoms: Be, Mg, Ca, Zn, Cd, and Hg for which a number of accurate calculations
is available.
We then apply the same approach to open-shell atoms, treating them in a
simplified manner similar to that of Ref.~\cite{DFG10}. Fractional
occupation numbers are used to rescale the terms containing contributions from
the open shells. The main advance of the present method compared with
Ref.~\cite{DFG10} is the all-order treatment of the positron-electron
interaction in the correlation potential. No further fitting is used or
needed. Remarkably, the results of the all-order calculations turn out to
be close to the estimates obtained in our previous work. These calculations
lend further support to the proposal to search for positron-bound states with
open-shell atoms through resonant annihilation or scattering. 

\section{Theory}

\subsection{Electron and positron basis sets}

The use of the coupled cluster technique requires construction of a
single-particle basis. For a positron interacting with an atom one
must have two separate basis sets -- one for electron states and one for
positron states. We use a standard B-spline technique in both
cases~\cite{Bspline}. The electron basis states are constructed by
diagonalizing the matrix of the relativistic Hartree-Fock (RHF)
Hamiltonian in the B-spline basis. The positron basis states are
constructed using the same set of B-splines and the RHF Hamiltonian in which
the sign of the direct potential is changed and the exchange potential is
omitted. Below we will use the following notation for the basis states:
indices $a,b,c$ refer to electron states in the core, indices $m,n,k,l$
refer to electron states above the core, indices $v,r,w$ refer to positron
states, and indices $i,j$ refer to any states. Numerical results
reported in Secs. \ref{sec:closed} and \ref{sec:open} were obtained using
the basis sets built from 30 B-splines of order 7 spanning the radial
coordinate from the origin to $R=30$~a.u., with angular momenta between
0 and 10.

\subsection{Singles-doubles equations}

The wave function of an atom with a positron in state $v$ can be
written in the SD approximation as an expansion
\begin{align}
|\Psi_{v}\rangle = &\left[ 1+\sum_{na}\rho_{na} a_n^\dagger a_a +
  \frac{1}{2}\sum_{mnab} \rho_{mnab} a_m^\dagger a^\dagger _n a_a a_b
   \right. \nonumber \\
&+\left. \sum_{r\neq v} p_{rv} a^\dagger _{r} a_{v}+ \sum_{rna}
  p_{rnva} a^\dagger_{r} a_{v} a_n^\dagger a_a\right]
|\Phi_{v}\rangle, \label{eq:psiv} 
\end{align}
where $|\Phi_{v}\rangle$ is the zeroth-order wave function of the
frozen-core atom in the relativistic Hartree-Fock approximation with
the positron in state $v$. It can be written as
\begin{equation}
|\Phi_{v}\rangle = a^\dagger_{v}|0_C\rangle,
\label{eq:0C}
\end{equation}
where $|0_C\rangle$ is the RHF wave function of the atomic core.

The expansion coefficients $\rho_{na}$ and $\rho_{mnab}$ in Eq.~(\ref{eq:psiv})
represent single- and double-electron excitations from the core. The
coefficients $p_{rv}$ represent excitations of the positron, and the
coefficients $p_{rnwa}$ represent simultaneous excitatons of the positron
and one of the electrons.

The SD equations for the core excitation coefficients do not depend on the
external particle and are the same in the electron and positron cases.
They are written as a set of equations for the single-excitation coefficients
$\rho_{ma}$ and double-excitation coefficients $\rho_{mnab}$
(see, e.g., Ref.~\cite{SD}),
\begin{equation}\label{eq:core1}
\begin{split}
(\epsilon_a-\epsilon_m)\rho_{ma} &= \sum_{bn}
\tilde{g}_{mban}\rho_{nb} \\
& +\sum_{bnk}g_{mbnk}\tilde{\rho}_{nkab}-\sum_{bcn}g_{bcan}\tilde{\rho}_{mnbc}\,, 
\end{split}
\end{equation}
and
\begin{equation}\label{eq:core2}
\begin{split}
&(\epsilon_a+\epsilon_b-\epsilon_m-\epsilon_n)\rho_{mnab} = g_{mnab}\\
&+\sum_{cd}g_{cdab}\rho_{mncd} + \sum_{kl}g_{mnkl}\rho_{klab}  \\
&+\sum_k g_{mnkb}\rho_{ka}-\sum_c g_{cnab}\rho_{mc} +
  \sum_{kc}\tilde{g}_{cnkb}\tilde{\rho}_{mkac} \\
&+\sum_k g_{nmka}\rho_{kb}-\sum_c g_{cmba}\rho_{nc} +
  \sum_{kc}\tilde{g}_{cmka}\tilde{\rho}_{nkbc} \,.
\end{split}
\end{equation}
In these equation $\tilde{g}_{mnkl} \equiv g_{mnkl} - g_{mnlk}$ and
$\tilde{\rho}_{mnkl} \equiv \rho_{mnkl} - \rho_{mnlk}$,
and the coefficients $g$ are the Coulomb integrals,
\begin{equation}\label{eq:g}
g_{mnkl}=\iint \psi_m^{\dag}(\mathbf{r}_1) \psi_n^{\dag}(\mathbf{r}_2)\frac{e^2}
{|\mathbf{r}_1 - \mathbf{r}_2|}\psi_k(\mathbf{r}_1)\psi_l(\mathbf{r}_2)
d\mathbf{r}_1d\mathbf{r}_2,
\end{equation}
involving electron states $\psi _k$, $\psi _l$, etc.

The core SD equations (\ref{eq:core1}) and (\ref{eq:core2}) are solved
iteratively until convergence is achieved. This is controlled by
the correlation correction to the energy of the core, 
\begin{equation}\label{eq:dec}
\delta E_C = \frac{1}{2}\sum_{mnab} g_{abmn}\tilde{\rho}_{nmba} .
\end{equation}

%----------------------------------------------------------------

After solving the SD equations for the core one can start iterating the SD
equations for the external particle. The SD equations for the positron can be
obtained by substituting the state $|\Psi_{v}\rangle$ from Eq.~(\ref{eq:psiv})
into the relativistic many-body Schr\"{o}dinger equation,
\begin{equation}\label{eq:S}
H|\Psi_{v}\rangle = \epsilon_0 |\Psi_{v}\rangle.
\end{equation}
Projecting the Schr\"{o}dinger equation onto $a^\dagger_w|0_C\rangle$
gives the equation for $p_{wv}$,
\begin{equation}\label{eq:v1}
(\epsilon_0-\epsilon_w)p_{wv} = 
 -\sum_{bm}q_{wbvm}\rho_{mb} + \sum_{bmr}q_{wbrm}p_{rmvb},
\end{equation}
Projecting Eq.~(\ref{eq:S}) onto $a^\dagger_{w}a^\dagger_{n}a_a|0_C\rangle$ gives
the equation for the double-excitation coefficient $p_{wnva}$,
\begin{equation}\label{eq:v2}
\begin{split}
(\epsilon_0+\epsilon_a -\epsilon_w-&\epsilon_n)p_{wnva} = q_{wnva} \\
-&\sum_{rm}q_{wnrm}p_{rmva} + \sum_m q_{wnvm}\rho_{ma}\\
-&\sum_b q_{wavb}\rho_{nb} + \sum_{mb}p_{wmvb}\tilde{g}_{mabn}\\
+&\sum_{rb}q_{warb}p_{rbvn} + \sum_{mb}q_{wmvb}\tilde{\rho}_{mabn}\, .
\end{split}
\end{equation}
Here $q_{wnva}$ is the Coulomb integral (\ref{eq:g}) involving positron states.

\begin{figure}
\centering
\epsfig{figure=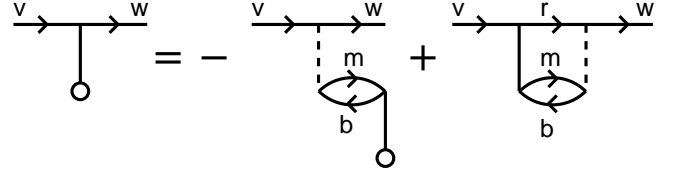,scale=0.7}
\caption{Diagrammatic form of the SD equation (\ref{eq:v1}) for the positron
single-excitation coefficient. Lines with arrows to the right are positron
and excited electron states, and those with arrows to the left are core
electron states (``holes''). The vertical solid line is the
double-excitation coefficient, and the solid line terminated by a circle
is the single-excitation coefficient. Dashed lines are the Coulomb
interactions.} 
\label{f:sd1}
\end{figure}

\begin{figure*}
\centering
\epsfig{figure=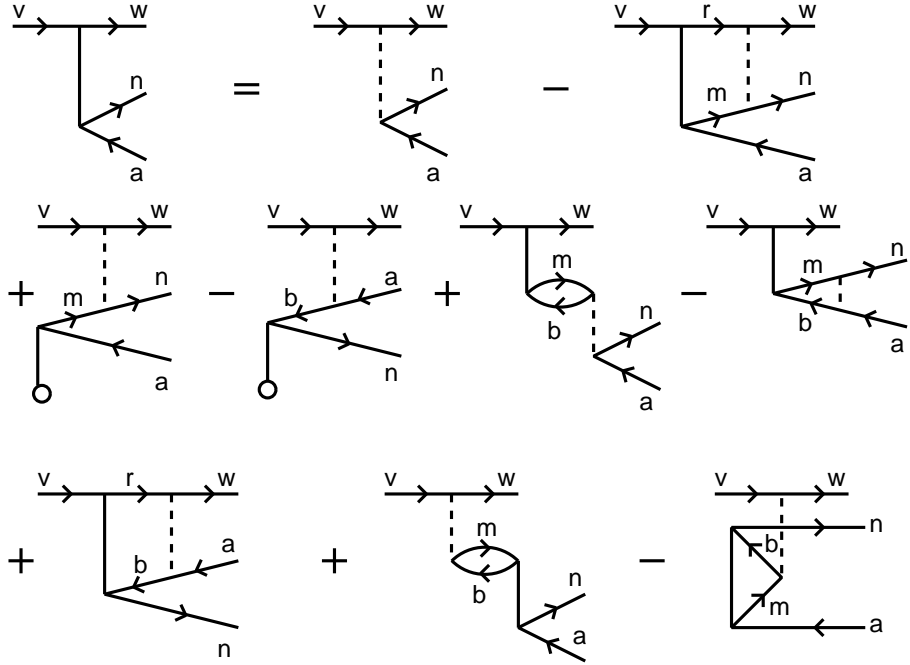,scale=0.7}
\caption{Diagrammatic form of the SD equation (\ref{eq:v2}) for the positron
double-excitation coefficient.} 
\label{f:sdp}
\end{figure*}

Equations (\ref{eq:v1}) and (\ref{eq:v2}) are presented graphically
in Figs. \ref{f:sd1} and \ref{f:sdp}. When solving these equations,
the correction to the energy of the positron state $v$,
\begin{equation}\label{eq:dev}
\delta \epsilon_v = - \sum_{mb} q_{vbvm}\rho_{mb} + \sum_{bmr}q_{vbrm}p_{rmvb},
\end{equation}
is used to control convergence.

%----------------------------------------------------------------------
%----------------------------------------------------------------------

%----------------------------------------------------------------------
%----------------------------------------------------------------------

\subsection{Positron bound state}

Apart from the absence of the exchange terms involving the positron, there is
another important difference between the SD equation (\ref{eq:v2}) for the
positron and those for external electrons in atoms or ions. In the latter
cases the SD equation for the double-excitation coefficient also contains
a term $\sum_r q_{wnra}p_{rv}$. The corresponding diagram is ``reducible'',
i.e., it contains two parts connected by the single line of the excited
valence particle. Including it ensures that the correction to the energy of
the valence state $v$ is determined in all orders.

When using the SD method for the positron, its eigenstate and energy eigenvalue
$\epsilon _0$ are obtained by matrix diagonalization (see below). The need for
this procedure
arises because of the absence of a good zeroth-order approximation for the
wave function of the bound positron. In the RHF approximation the positron-atom
interaction is repulsive, and all of the single-particle positron
basis states lie in the continuum. On the other hand, the B-spline basis is
effectively
complete for the positron bound state to be obtained by matrix diagonalization.
To do this, the SD equations (\ref{eq:v1}) and (\ref{eq:v2}) should be iterated
for every positron basis state used to construct effective single-particle
Hamiltonian matrix. A similar situation occurs for negative
ions~\cite{DG94,DFGH12} and when combining the SD method with the configuration
interaction technique to obtain many-electron wave functions~\cite{Jiang}.
In all of these cases the term $\sum_r q_{wnra}p_{rv}$ should be removed from
the SD equations, as it is taken into account via matrix diagonalization.

The wavefunction of the bound positron is found as an expansion over the
positron basis states,
\begin{equation}
\psi_p = \sum_v c_v \psi_v .
\label{eq:psip}
\end{equation}
Equations (\ref{eq:v1}) and (\ref{eq:v2}) are iterated for every basis
state $v$ in the expansion (\ref{eq:psip}). The energy parameter
$\epsilon_0$ in these equations is the (unknown) energy of the positron
state (\ref{eq:psip}). The energy $\epsilon_0$ and the expansion
coefficients $c_v$ are found by solving the eigenvalue problem
\begin{equation}
\hat \Sigma X = \epsilon_0 X,
\label{eq:sx}
\end{equation}
where $X$ is the vector of expansion coefficients $c_v$, $\epsilon _0$ is the
lowest eigenvalue (which must be negative), and the elements of the effective
Hamiltonian matrix $\hat \Sigma$ are given by
\begin{equation}\label{eq:sigma}
\sigma_{vw} = \epsilon_v\delta_{vw} - \sum_{mb} q_{wbvm}\rho_{mb} +
\sum_{bmr}q_{wbrm}p_{rmvb} .
\end{equation}
The first term on the righ-hand side of Eq.~(\ref{eq:sigma}) represents
the positron energies in the static RHF approximation. The second and third
terms describe the effect of electron-positron correlations. These two
terms are the positron-atom \textit{self-energy}, which is given by
right-hand side of the diagrammatic equation in Fig.~\ref{f:sd1}.

Since the SD equations (\ref{eq:v1}) and (\ref{eq:v2}) depend on the
energy $\epsilon_0$ which is found later from Eq.~(\ref{eq:sx}), we start
with an initial guess for $\epsilon_0$. The calculations are then performed
iteratively, solving the SD equations (\ref{eq:v1}) and (\ref{eq:v2}) and
diagonalizing the matrix (\ref{eq:sigma}) several times until $\epsilon _0$
has converged. In practice this takes about 5 global iterations.

\subsection{Third-order corrections}

The SD equations account for the second-order contribution
and selected classes of higher-order correlation diagrams in all orders. In
particular, the electron-positron ladder diagram series, which describes
virtual Ps formation, is included in full. However, beginning with the
third order, the SD approximation misses certain terms. It is well known that
the missing third-order terms can give sizeable corrections
to the energy in atomic systems (see, e.g., Ref.~\cite{E3extra}).
Including these terms can lead to significant improvements in the accuracy
of the results, and we include these contributions for the positron
bound states with atoms.

\begin{figure*}[t!]
\centering
\epsfig{figure=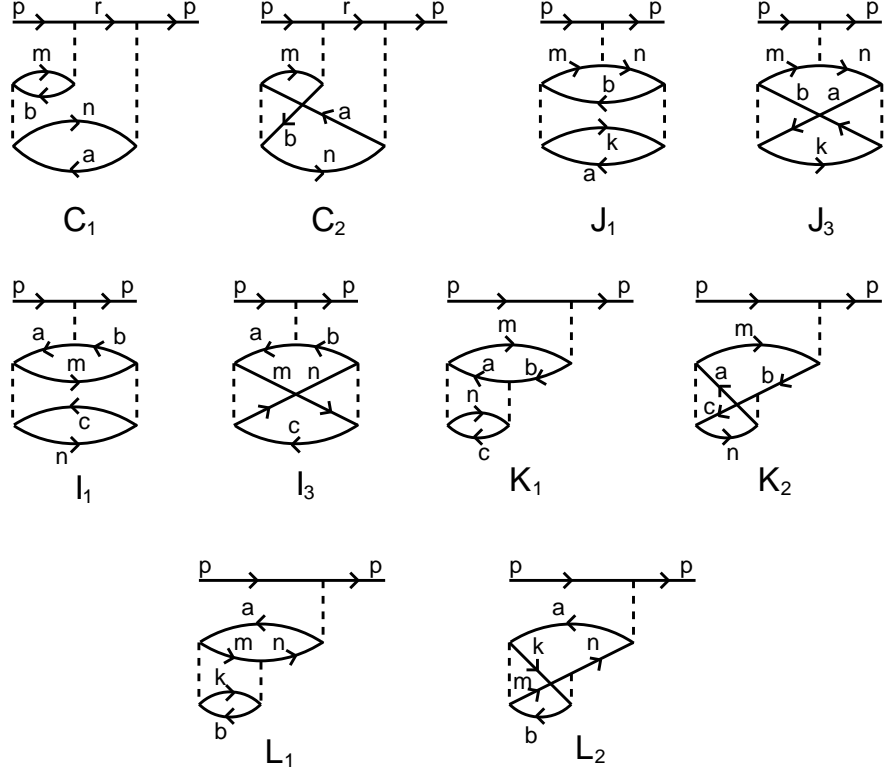,scale=0.65}
\caption{Third-order positron-atom interaction diagrams not
included into the SD equations. Notation of Ref.~\cite{E3} is used.} 
\label{f:e3}
\end{figure*}

The third-order terms not included in the SD approximation are
listed in Ref.~\cite{E3extra} for the case of electron interacting
with a closed-shell ion. Similar terms for the positron case can be
obtained by changing the sign of Coulomb integrals involving the positron
and removing exchange. All resulting third-order diagrams are shown in
Fig.~\ref{f:e3}. The corresponding perturbation-theory corrections to
the energy of the positron state $p$ are
\begin{eqnarray}
E^{(3)}_C &=& \sum_{abmnr}
\frac{q_{pmrb}q_{rnpa}\tilde{g}_{bamn}}
{(\epsilon_{0a} -\epsilon_{rn})(\epsilon_{ab} - \epsilon_{mn})}, \label{eq:C}\\
E^{(3)}_I &=& \sum_{abcmn}
\frac{q_{papb}g_{acmn}\tilde{g}_{mnbc}}
{(\epsilon_{bc} -\epsilon_{mn})(\epsilon_{ab} - \epsilon_{mn})}, \label{eq:I}\\
E^{(3)}_J &=-& \sum_{abmnk}
\frac{q_{pmpn}g_{bamk}\tilde{g}_{bank}}
{(\epsilon_{ab} -\epsilon_{mk})(\epsilon_{ab} - \epsilon_{nk})}, \label{eq:J}\\
E^{(3)}_K &=& \sum_{abcmn}
\frac{q_{pmpb}g_{acmn}\tilde{g}_{acnb}}
{(\epsilon_{ac} -\epsilon_{nm})(\epsilon_b - \epsilon_m)}, \label{eq:K} \\
E^{(3)}_L &=-& \sum_{abmnk}
\frac{q_{papn}g_{abmk}\tilde{g}_{mbnk}}
{(\epsilon_{ab} -\epsilon_{mk})(\epsilon_a - \epsilon_n)}.
\label{eq:L}
\end{eqnarray}
Here $\epsilon_{ij} = \epsilon_i + \epsilon_j$, $\epsilon_0$ is the
positron energy found by solving the eigenvalue problem (\ref{eq:sx}), and
the positron state $p$ is given by Eq.~(\ref{eq:psip}).

\section{Closed-shell atoms}\label{sec:closed}

For a positron interacting with a closed-shell atom all sums over the
projections of angular momenta of electron and positron states in the
SD equations (\ref{eq:core1}), (\ref{eq:core2}), (\ref{eq:v1}), and
(\ref{eq:v2}) and third-order corrections (\ref{eq:C})--(\ref{eq:L}), are done
analytically, together with the angular reduction of the Coulomb integrals.
Such systems are easiest from the computational point of view. There are also
a number of other calculations of positron binding of different level of
sophistication for several such atoms. We use these systems to test our
approach. The results of calculations of positron $s$-wave bound states with
Be, Mg, Ca, Zn, Cd, and Hg are shown in Table~\ref{t:1}. These six atoms share
the following features. First, they have closed electronic shells. Secondly, the
dipole polarizabilities $\alpha _d$ of these atoms are sufficiently large to
provide strong attraction and ensure positron binding.
Finally, the ionization potentials $I$ of these atoms (with the exception of
calcium) are greater
than the Ps binding energy (6.8~eV). Hence, the Ps-formation
channel is closed at low positron energies, and the positron-atom bound states
are stable against dissociation into positive ion $+$ Ps.

\begin{table*}
\caption{Positron binding energies (BE, in meV) for
closed-shell atoms obtained using the SD equations (SD) with third-order
correction (E3), and by other methods, with best available predictions in bold.}
\label{t:1}
\begin{ruledtabular}
\begin{tabular}{rc rrrrr l}
\multicolumn{2}{c}{Atom} &
\multicolumn{1}{c}{$I$} &
\multicolumn{1}{c}{$\alpha_d$\footnotemark[1]} &
\multicolumn{3}{c}{BE, this work} &
\multicolumn{1}{c}{BE from other calculations} \\
\multicolumn{1}{c}{$Z$} & &
\multicolumn{1}{c}{(eV)} &
\multicolumn{1}{c}{(a.u.)} &
\multicolumn{1}{c}{SD} &
\multicolumn{1}{c}{E3} &
\multicolumn{1}{c}{Total} & \\
\hline
 4 & Be & 9.32 & 38 & 157 &  26 & 184 & 45.90~\cite{Ryzhikh98},
 {\bf 85.63}~\cite{Mitroy01}, {\bf 83.8}~\cite{Bromley02}, 3.0~\cite{Szmytkowski-a},
 33~\cite{Mella02}   \\
12 & Mg & 7.64 & 72 & 475 &  37 & 512 & 870~\cite{Dzuba95},
425~\cite{Mitroy01}, {\bf 439}~\cite{Bromley02a}, 985~\cite{King96},
15~\cite{Szmytkowski-a}, 457~\cite{Mella02}, 125~\cite{McEachran} \\
20 & Ca & 6.11 & 154 & 1114 &  50 & 1164 & {\bf 1139}\footnotemark[2]~\cite{Bromley02a} \\
30 & Zn & 9.39 & 42 & 143 &  40 & 183 & 230~\cite{Dzuba95}, 39~\cite{Mitroy99},
{\bf 103}~\cite{Bromley02b}, 0.01~\cite{Szmytkowski-b}, 53~\cite{McEachran}  \\
48 & Cd & 8.99 & 49 & 204 & 70 & 274 & 350~\cite{Dzuba95}, {\bf 166}~\cite{Bromley02b}, 1.5~\cite{Szmytkowski-b} \\
80 & Hg & 10.43 & 38 & 59 & 53 & 112 & {\bf 45}~\cite{Dzuba95}      \\
\end{tabular}
\footnotetext[1]{Static dipole polarizabilities from Ref. \cite{Miller}.}
\footnotetext[5]{Positron-atom BE obtained from the calculated BE of 449 meV
with respect to the Ca$^++$Ps threshold.}
\end{ruledtabular}
\end{table*}

The SD column of Table~\ref{t:1} shows the binding energies (BE) $|\epsilon_0|$
(in meV) found by solving the eigenvalue problem (\ref{eq:sx}). The E3
column gives the third-order corrections (\ref{eq:C})--(\ref{eq:L}).
The next column is the total obtained by adding the third-order correction
to the BE from the SD calculations. We also examined the effect of finite box
size on the binding energy~\cite{DFG99}, and found it negligible.
The rest of the Table shows the results of earlier calculations. The most
accurate among them are probably the calculations by the Mitroy group
\cite{Ryzhikh98,Mitroy01,Bromley02,Bromley02a,Bromley02b}.
Our results are generally slightly larger, but within 100 meV of the best
earlier predictions. Adding third-order corrections increases the BE
in all the systems, and in most cases increases the difference between the
present and best earlier results.

Note the scatter of the results between different groups and even between
different calculations by the same group. This reflects the fact that the
positron-atom binding energy is very sensitive to the correlations. In the
static field RHF approximation the positron is repelled from the
neutral atom. The binding is solely due to correlations. Most of the
correlation energy is required to get the system across the threshold
for binding, while the final binding energy is a result of a relatively
small ``surplus'' of the correlation energy. Note that according to our calculations, positrons do not bind to either Pd or Pt in the atomic ground states. However, both atoms bind in excited states.

\section{Open-shell atoms}\label{sec:open}

In this section we consider open-shell atoms which were suggested in
our previous work~\cite{DFG10} as good candidates for experimental
detection of positron-atom bound states via resonant annihilation or
scattering. We treat open-shell systems in an approximate way, by introducing
fractional occupation numbers. For example, the ground-state electron
configuration of neutral iron is $3d^64s^2$ above the Ar-like core.
We treat it as a closed-shell system but reduce the contribution of the $3d$
subshell to the correlation potential by the factor 0.6.
%This includes the Hartree-Fock equations, the SD equations
%(\ref{eq:core1},\ref{eq:core2},\ref{eq:v1},\ref{eq:v2}) and the
%third-order energies
%(\ref{eq:C},\ref{eq:I},\ref{eq:J},\ref{eq:K},\ref{eq:L}). 
Both members of the fine-structure multiplet, $3d_{3/2}$ and $3d_{5/2}$,
are treated identically and the corresponding terms are rescaled by the same
factor 0.6.

Note, however, that the SD equations (\ref{eq:v2}) are left unchanged. There
is strong cancellation between different terms in these equations and
rescaling of the core contribution is an insufficiently
accurate procedure. It breaks the delicate balance between different
terms, leading to unreliable results. The rescaling is done when constructing
the correlation potential matrix $\sigma_{vw}$, by reducing the terms
corresponding to the open subshells when summing over the hole states $b$ in
Eq.~(\ref{eq:sigma}). The fact that this procedure works well is supported
by good agreement between the present calculations of positron binding to
copper with our previous calculations (see Table \ref{t:2}).

\begin{table*}
\caption{Positron binding energies (BE, in eV) for
open-shell atoms obtained using the SD equations (SD) with third-order
correction (E3). In this table $I$ is the ionization energy from the lowest state
of a given configuration. The last collumn shows semiempirical
values from Ref.\cite{Babikov}.}
\label{t:2}
\begin{ruledtabular}
\begin{tabular}{rll rrrrr rr}
\multicolumn{2}{c}{Atom} & 
\multicolumn{1}{c}{Valence} &
\multicolumn{1}{c}{$I$} &
\multicolumn{1}{c}{$\alpha_d$\footnotemark[1]} &
\multicolumn{3}{c}{BE, this work} &
\multicolumn{2}{c}{BE, other calculations}  \\
\multicolumn{1}{c}{$Z$} & &
\multicolumn{1}{c}{configuration} &
\multicolumn{1}{c}{(eV)} &
\multicolumn{1}{c}{(a.u.)} &
\multicolumn{1}{c}{SD} &
\multicolumn{1}{c}{E3} &
\multicolumn{1}{c}{Total} & 
\multicolumn{1}{c}{Ref.\cite{DFG10}} &
\multicolumn{1}{c}{Ref.\cite{Babikov}} \\
\hline
21 & Sc & $3d 4s^2$   & 6.56 & 120 & 0.908 & 0.129 & 1.037 &  & 0.75(2) \\ 
   &    & $3d^2 4s$   & 5.13 &     & 0.849 & 0.109 & 0.958 &  &  \\

22 & Ti & $3d^2 4s^2$ & 6.83 & 98.5 & 0.785 & 0.110  & 0.896 &  & 0.84(3) \\ 
   &    & $3d^3 4s$   & 6.02 &      & 0.727 & 0.097  & 0.825 &  &  \\

23 &  V & $3d^3 4s^2$ & 6.74 & 83.7 & 0.678 & 0.097  & 0.775 &  & 0.81(3) \\ 
   &    & $3d^4 4s$   & 6.48 &      & 0.602 & 0.086 & 0.689 &  &  \\

24 & Cr & $3d^5 4s$ & 6.76 & 78.3 & 0.488 & 0.077  & 0.565 &  & 0.54(8) \\ 
   &    & $3d^4 4s^2$   & 5.80 &      & 0.590 & 0.086 & 0.676 &  &  \\

25 & Mn & $3d^5 4s^2$ & 7.43 & 63 & 0.496 & 0.077  & 0.574 &  & 0.53(3) \\ 
   &    & $3d^6 4s$   & 5.32 &    & 0.382 & 0.068 & 0.450 &  &  \\

26 & Fe & $3d^6 4s^2$ & 7.90 & 56.7 & 0.429 & 0.069 & 0.498 & 0.28 & 0.37(3) \\ 
   &    & $3d^7 4s$   & 7.04 &      & 0.315 & 0.061 & 0.376 & 0.09 & \\ 

27 & Co & $3d^7 4s^2$ & 7.88 & 50.7 & 0.360 & 0.061 & 0.422 & 0.26 & 0.36(3) \\
   &    & $3d^8 4s$   & 7.45 &      & 0.243 & 0.053 & 0.297 & 0.08 & \\

28 & Ni & $3d^8 4s^2$ & 7.64 & 45.9 & 0.295 & 0.055 & 0.350 & 0.24 & 0.42(3) \\
   &    & $3d^9 4s$   & 7.61 &      & 0.173 & 0.046 & 0.220 & 0.07 &    \\

29 & Cu &   $3d^{10} 4s$ & 7.72 & 41 & 0.125 & 0.040 & 0.166 &
0.17(2)\footnotemark[2] & 0.152(2)\footnotemark[3]    \\

39 &  Y & $4d 5s^2$   & 6.22 & 153 & 0.845 & 0.256 & 1.102 &  & 0.6(19) \\ 
   &    & $4d^2 5s$   & 4.86 &     & 0.683 & 0.258 & 0.942 &  & \\

40 & Zr & $4d^2 5s^2$ & 6.63 & 121 & 0.729 & 0.209 & 0.939 &  & 0.8(11) \\ 
   &    & $4d^3 5s$   & 6.12 &    &  0.623 & 0.208 & 0.831 &  & \\

41 & Nb & $4d^4 5s$ & 6.76 & 106 & 0.527 & 0.172 & 0.699 &  & 0.6(14) \\ 
   &    & $4d^3 5s^2$ & 6.62 &   & 0.658 & 0.178 & 0.836 &  &  \\

42 & Mo & $4d^5 5s$ & 7.09 & 86.4 & 0.442 & 0.145 & 0.587 &  & 0.45(4) \\ 
   &    & $4d^4 5s^2$ & 5.73 &    & 0.583 & 0.155 & 0.739 &  &  \\

43 & Tc & $4d^5 5s^2$ & 7.28 & 76.9 & 0.461 & 0.133 & 0.594 & 0.46 & 0.62(4) \\ 
   &    & $4d^6 5s$   & 6.96 &      & 0.355 & 0.124 & 0.479 & 0.23 &  \\

44 & Ru & $4d^7 5s$ & 7.36 & 65 & 0.310 & 0.109  & 0.419 & 0.21 & 0.30(95) \\ 
   &    & $4d^6 5s^2$ & 6.43 &  & 0.461 & 0.121  & 0.583 &      & \\ 

45 & Rh & $4d^8 5s$ & 7.46 & 58 & 0.260 & 0.095  & 0.355 & 0.20 & 0.24(86) \\
%   &    & $4d^9 $   & 7.05 &   & -0.015 & 0.011  & $<0$ & 0.10 & \\

46 & Pd & $4d^{10}$ & 8.34 & 32 & $<0$ & $-$ & $<0$  &  & $<0$ \\
   &    & $4d^9 5s$ & 7.52 &   & 0.205 & 0.083 & 0.288 &     &    \\
   &    & $4d^8 5s^2$ & 5.23 &   & 0.361 & 0.097 & 0.459 &     &    \\

71 & Lu & $5d 6s^2$   & 5.43 & 148 & 0.188 & 0.221 & 0.410 &  & 0.2(11) \\ 
%   &    & $5d^2 6s$   & 3.09 &     & 2.559 & 0.187 & 2.746 &  &  \\

72 & Hf & $5d^2 6s^2$ & 6.83 & 109 & 0.305 & 0.198 & 0.503 &  & 0.8(10) \\ 
   &    & $5d^3 6s$   & 5.08 &     & 0.349 & 0.190 & 0.539 &  &  \\

73 & Ta & $5d^3 6s^2$ & 7.55 & 88.4 & 0.274 & 0.166  & 0.441  & 0.45 & 0.6(8) \\ 
   &    & $5d^4 6s $  & 6.34 &  & 0.126 & 0.147 & 0.273 &  &  \\

74 &  W & $5d^4 6s^2$ & 7.86 & 74.9 & 0.235 & 0.141  & 0.377 & 0.46 & 0.4(7) \\ 
   &    & $5d^5 6s$   & 7.50 &  & 0.083 & 0.118 & 0.201 & 0.30 &  \\

75 & Re & $5d^5 6s^2$ & 7.83 & 65 & 0.202 & 0.121 & 0.324 &  & 0.42(12) \\ 
   &    & $5d^6 6s$   & 6.38 &    & 0.051 & 0.096 & 0.147 &  &  \\

76 & Os & $5d^6 6s^2$ & 8.44 & 57 & 0.167 & 0.105 & 0.273 & 0.47 & 0.3(5) \\ 
   &    & $5d^7 6s$   & 7.80 &   & 0.025 & 0.079 & 0.105 & 0.29 & \\ 

77 & Ir & $5d^7 6s^2$ & 8.97 & 51 & 0.137 & 0.091 & 0.229 & 0.46 & 0.2(4) \\
   &    & $5d^8 6s$   & 8.62 &   & 0.005 & 0.067 & 0.072 & 0.28 & \\

78 & Pt & $5d^9 6s$ & 8.96 & 44 & $<0$ & $-$ & $<0$
% 0.055\footnotemark[4] 
&  0.27 & $<0$ \\
   &    & $5d^8 6s^2$ & 8.86 &   & 0.111 & 0.080 & 0.191 & 0.46 &    \\
   &    & $5d^{10}$   & 8.20 &   & 0.020 & 0.023 & 0.044 & 0.23 &    \\
\end{tabular}
\footnotetext[1]{Ground-state atomic static dipole polarizabilities
from Ref.~\cite{Miller}.}
\footnotetext[2]{Configuration-interaction with many-body theory,
Ref.~\cite{DFG99}.}
\footnotetext[3]{Stochastic variation method, Ref.~\cite{MBR02}.}
%\footnotetext[4]{Correction due to the finite box size $\Delta E$ = 10 meV
%is also included.}
\end{ruledtabular}
\end{table*}

%To stay close to the original equations and avoid large computational
%error we consider only systems were occupation of subshells is close
%to 100\%. In other words, we only consider atoms with almost filled
%outermost $nd$ subshell while all other subshells are fully filled.

This approach does not distinguish between different states of the
same configuration. This means that if the positron is found to be bound
to an atom in a given electronic configuration, it is predicted to be
bound to all states of this configuration. This should be expected at least
for the states where the excitation energy is smaller than the binding energy.
However, this may still be true for higher-lying states. The ability of an
atom in an excited state to bind the positron is what is needed for the
resonant process proposed in Ref.~\cite{DFG10}. As it was argued in
Ref.~\cite{DFG10} and as we will see below, there are many such atoms.

Table \ref{t:2} shows the results of our calculations of positron binding
energies for a number of atoms with an open $d$-shell. These values are
compared with the binding energies calculated in Ref.~\cite{DFG10} and
with semiempirical estimates of Ref.~\cite{Babikov}. Within the accuracy of
the current approach all results appear to be in good accord with
each other. There is a correlation between the occupation of
the $d$-shell and the binding energies. In most cases the closer the
occupation is to 100\% the smaller is the binding energy, and the smaller
are the differences between the values obtained using different approaches.
In the end, the results of the present work confirm the claim of
Ref.~\cite{DFG10} that many open-shell atoms do bind the positron not only in
the ground state but also in excited states.
Note that according to our calculations, positrons do not bind to either Pd
or Pt in their ground states. However, both atoms bind in excited
states.

The final energies of positron bound states and resonances are presented in
Table~\ref{t:3}. The resonant energy is the energy of the incident positron
$ \varepsilon$ for which capture into a bound state with an excited atom is
possible. The resonant and binding energies are related to each other
by
\begin{equation}\label{eq:er}
\varepsilon_r = E_{\rm ex} - \varepsilon_b,
\end{equation}
where $E_{\rm ex}$ is the atomic excitation energy measured with respect to
the ground state. To observe the resonance, one requires $ \varepsilon_r>0$,
which means that the excitation energy must be greater than binding energy
($E_{\rm ex}> \varepsilon_b$). In lighter atoms, the fine-structure splitting
of the ground state is too small to satisfy this condition. However, in the
heavier atoms, the fine structure can be larger than $ \varepsilon_b$, and the
corresponding low-lying excited states can form resonances. There is also
another condition for observing narrow resonances. The resonance energy should
be smaller than the Ps formation threshold, $\varepsilon_r < I-6.8 $.
For resonances lying above this energy, the Ps formation channel will be open.
This can reduce the resonance lifetimes and make the resonances too broad to
be observed. Also, at these energies the positron-atom annihilation signal due
to resonances will be ``drowned'' by the annihilation within the positronium.
Table~\ref{t:3} shows resonances which satisfy or approximately satisfy
these two conditions. Note that we also show some states with negative
$ \varepsilon_r $ values, which correspond to weakly bound positron states.
Given the uncertainty in our calculation, they may in fact turn out to
be low-lying resonances.

\begin{table*}
\caption{Final binding and resonant energies ($ \varepsilon_b$ and
$\varepsilon_r$, in eV) for the detection of positron-atom bound states through
resonant annihilation or scattering.}
%$E_r = E_{ex} - E_b$}
\label{t:3}
\begin{ruledtabular}
\begin{tabular}{rcl llll}
\multicolumn{2}{c}{Atom} & 
\multicolumn{1}{c}{Valence} &
\multicolumn{1}{c}{Excited states} &
\multicolumn{1}{c}{$E_{\rm ex}$\footnotemark[1]} &
\multicolumn{1}{c}{$ \varepsilon_{b}$} &
\multicolumn{1}{c}{$ \varepsilon_{r}$\footnotemark[2]} \\
\multicolumn{1}{c}{$Z$} & &
\multicolumn{1}{c}{configuration} &&&& \\
\hline
22 & Ti & $3d^2 4s^2$ & $^1$D$_{2}$ & 0.90 & 0.90 & 0.0 \\
   &    & $3d^3 4s $ & $^5$F$_{1},^5$F$_{2}$ & 0.813, 0.818 & 0.825 & $-0.012$,
$-0.007$ \\
      
27 & Co & $3d^8 4s$   & $^4$F$_{9/2}, ^4$F$_{7/2}, ^4$F$_{5/2}, ^4$F$_{3/2}$ & 0.43, 0.51, 0.58, 0.63 & 0.30 & 0.13, 0.21, 0.28, 0.33 \\
   &    &             & $^2$F$_{7/2}$ & 0.92 & 0.30 & 0.62 \\
   
28 & Ni & $3d^9 4s$   & $^1$D$_{2}$ & 0.42 & 0.22 &  0.20  \\

43 & Tc & $4d^6 5s$   & $^6$D$_{5/2},^6$D$_{3/2},^6$D$_{1/2}$ & 0.46, 0.50, 0.52 & 0.48 & $-0.02$, 0.02, 0.04 \\

44 & Ru & $4d^7 5s$   & $^5$F$_{1}$ & 0.39 & 0.42 & $-0.03$ \\
   &    &             & $^3$F$_{4},^3$F$_{3}$ & 0.81, 1.00 & 0.42 & 0.39, 0.58 \\
   &    &             & $^5$P$_{2}$ & 1.00 & 0.42 & 0.58 \\
 
45 & Rh & $4d^8 5s$   & $^4$F$_{5/2},^4$F$_{3/2}$ & 0.322, 0.43 & 0.355 & $-0.033$, 0.08 \\ 
   &    &             & $^2$F$_{7/2},^2$F$_{5/2}$ & 0.71, 0.96 & 0.36 & 0.35, 0.60 \\

46 & Pd & $4d^9 5s$   & $^2$[5/2]$_{3},^2$[5/2]$_{2}$ & 0.81, 0.96 & 0.29 & 0.52, 0.67 \\
   
72 & Hf & $5d^2 6s^2$ & $^3$F$_{4}$ & 0.57 & 0.50 & 0.07 \\

73 & Ta & $5d^3 6s^2$ & $^4$F$_{7/2}, ^4$F$_{9/2}$ & 0.49, 0.70 & 0.44 & 0.05, 0.26 \\   
   &    &             & $^4$P$_{1/2},^4$P$_{3/2},^4$P$_{5/2}$ & 0.75, 0.75, 1.15 & 0.44 & 0.31, 0.31, 0.71 \\
   &    &             & $^2$G$_{7/2}$ & 1.20 & 0.44 & 0.76 \\
   
74 & W  & $5d^4 6s^2$ & $^5$D$_{2},^5$D$_{3}$ & 0.41, 0.60 & 0.38 & 0.03, 0.22 \\         
   &    &             & $^4$D$_{3}$ & 0.77 & 0.38 & 0.39 \\
   &    &             & $^3$P2$_{0}$ & 1.18 & 0.38 & 0.80 \\
   &    & $5d^5 6s$   & $^7$S$_{3}$ & 0.36 & 0.20 & 0.16 \\

75 & Re  & $5d^5 6s^2$ & $^4$P$_{5/2}$ & 1.44 & 0.32 & 1.12 \\

76 & Os & $5d^6 6s^2$ & $^5$D$_{2},^5$D$_{3},^5$D$_{1},^5$D$_{0}$ & 0.34, 0.52, 0.71, 0.76 & 0.27 & 0.07, 0.25, 0.44, 0.49 \\

   &    &             & $^3$H$_{5},^3$H$_{4}, ^3$H$_{6}$ & 1.78, 1.84, 1.84 & 0.27 & 1.51, 1.57, 1.57 \\

   &    & $5d^7 6s$   & $^5$F$_{5},^5$F$_{4}$ & 0.64, 1.08 & 0.10 & 0.54, 0.98 \\

77 & Ir & $5d^7 6s^2$ & $^4$F$_{3/2},^4$F$_{5/2},^4$F$_{7/2}$ & 0.51, 0.72, 0.78 & 0.23 & 0.28, 0.49, 0.55 \\

   &    &             & $^2$G$_{9/2},^2$G$_{7/2}$ & 1.73, 2.20 & 0.23 & 1.50, 1.97 \\
   &    &             & $^4$P$_{5/2}, ^4$P$_{3/2}, ^4$P$_{1/2}$ & 2.00, 2.30, 2.51 & 0.23 & 1.77, 2.07, 2.27 \\

   &    &             & $^2$H$_{11/2}$ & 2.43 & 0.23 & 2.20 \\
   &    & $5d^8 6s$   & $^4$F$_{9/2},^4$F$_{7/2},^4$F$_{5/2},^4$F$_{3/2}$ & 0.35, 0.88, 1.22, 1.47 & 0.07 & 0.28, 0.81, 1.15, 1.40 \\

   &    &             & $^2$P$_{3/2}, ^2$P$_{1/2}$ & 1.31, 1.55 & 0.07 & 1.24, 1.48 \\
   
   &    &             & $^2$F$_{5/2},^2$F$_{7/2}$ & 1.51, 1.62 & 0.07 & 1.44, 1.55 \\
   &    &             & $^4$P$_{5/2}$ & 1.60 & 0.07 & 1.53 \\

%78 & Pt & $5d^9 6s$   & $^3$D$_{2}, ^3$D$_{1}$ & 0.10, 1.25 & 0.06 & 0.04, %1.19 \\      
%   &    &             & $^1$D$_{2}$ & 1.67 & 0.06 & 1.61 \\   
78 & Pt & $5d^8 6s^2$ & $^3$F$_{3}, ^3$F$_{2}$ & 1.25, 1.92 & 0.19 & 1.06, 1.73 \\
   &    &             & $^3$P$_{2}$ & 0.81 & 0.19 & 0.62 \\
   &    & $5d^{10}$   & $^1$S$_{0}$ & 0.76 & 0.04 & 0.72 \\
\end{tabular}
\footnotetext[1]{Atomic excitation energy with respect to
 the ground state from Ref.~\cite{NIST}.}
\footnotetext[2]{Resonance energy from Eq.~(\ref{eq:er}).}
\end{ruledtabular}
\end{table*}

\section{Conclusion}

The coupled cluster single-double approach has been used to calculate
positron binding energy for a number of open-shell atoms. The binding
energies are in good agreement with previous estimates and indicate that atoms
with open $d$-shells can bind positron not only in their ground but also in
excited states. Many of the atoms considered appear to be good candidates for
studying positron binding to atoms through resonant annihilation or resonant
scattering. 

\begin{acknowledgments}
This work was funded in part by the Australian Research Council. GG is grateful
to the Gordon Godfrey fund (UNSW) for support.
\end{acknowledgments}

\end{document}